# Self-supervised Enhanced Radar Imaging Based on Deep-Learning-Assisted Compressed Sensing

Shaoyin Huang[1]

[1]School of computer science, Sun Yat-sen University, China

huangshy95@mail2.sysu.edu.cn

**ABSTRACT**

Traditional radar imaging methods suffer from the problems of low resolution and poor noise suppression. We propose a new radar imaging method based on Self-supervised deep-learning-assisted compressed sensing (SS-DL-CS-Net). The original radar image is used as the inputs of network. The network is trained to learn the mapping function between the original radar image and the high quality radar image. However, the high quality radar image can't be obtained. We solve this problem by used the sparsity of radar image. The original radar image and image with the zeros value is used as the reference of network. Ours network don't need data in advance to train. Real radar data are used to evaluate the performance of the proposed method. The experimental results demonstrate the superiority of the proposed method

**Index Terms**-Self-supervised, deep learning, compressed sensing, radar imaging

## 1. INTRODUCTION

Radar image is widely used in civil and military fields for long-distance, all-day, and all-weather observations [1]. However, traditional radar imaging methods suffer from the problems of low resolution and poor noise suppression. Super-resolution of radar imaging with effective denoising can be achieved by sparse representation [2], [3]. Additional target structure can be further taken into consideration in sparse representation for an improved target feature preservation [4], [5]. However, the computational complexity of sparse representation is high. Recently, the convolution neural network is applied to radar imaging [6-8]. Due to the fast and effective performance of convolutional neural network, many research had paid attention to it. However, the training of convolutional neural network requires a large number of data sets. It is hard to obtain the effective training set in reality. Many researchers generate training sets by simulation. However, the training sets generated by simulation cannot correspond well to the real images. The high quality radar image cannot be obtained. Inspired by the Self-supervised [9], deep-learning-assisted compressed sensing [10], [11], we propose a new radar imaging method based on Self-supervised deep-learning-assisted compressed sensing(SS-DL-CS-net).

## 2. METHOD

A. Problem Analysis

A basic linear inverse problem as can be considered as:

$$y = Ax + n$$

where $y \in R^m$, $A \in R^{m \times n}$ and $x \in R^n$ are known. $n$ is an unknown noise

$$x = x' + n1$$

where $n1$ is an unknown noise, $x'$ is original signal [8].

$$y' = Ax'$$

where $y'$ is the "true" and unknown image to be estimated [8]. To obtain the $y'$, this problem can be represented as:

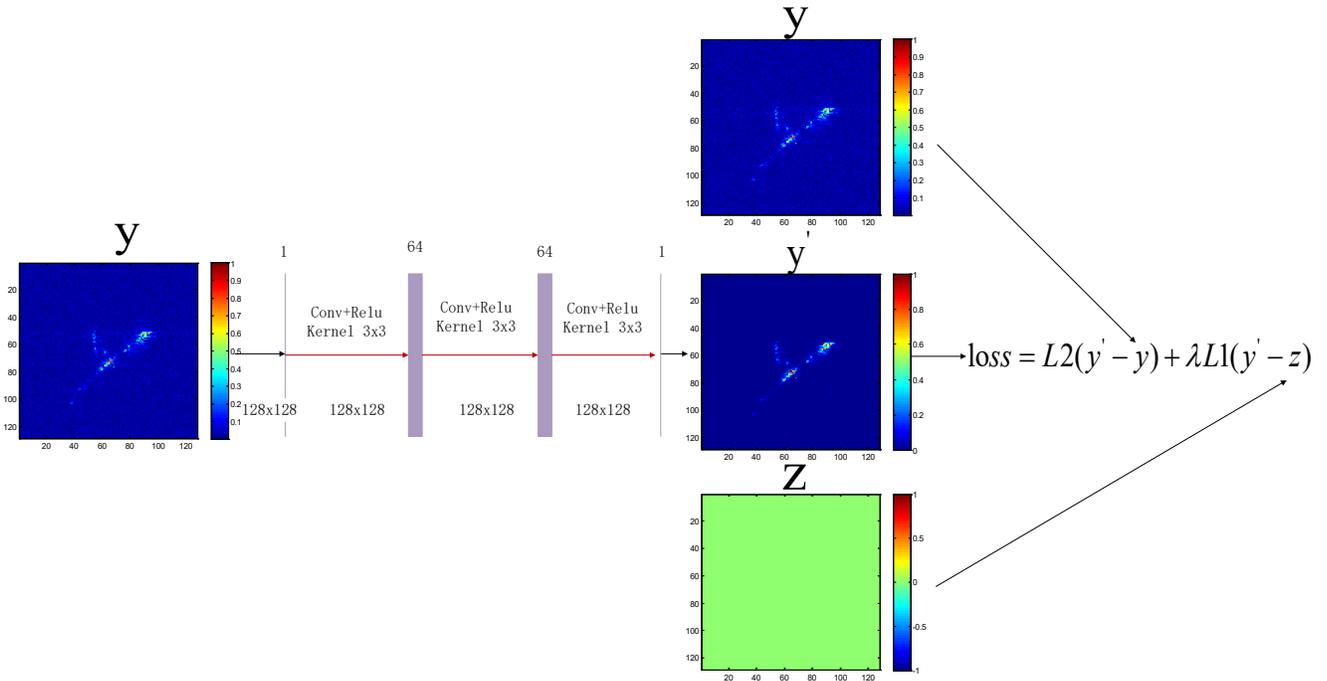

Fig.1 The structure of CNN named SS-DL-CS-Net

$$x' = \operatorname*{argmin}_{x} ||Ax - y||^2 + \lambda ||Ax||$$

the paper [10] propose a method to learn the mapping function between the $x$ and $x'$. It can be represented as:

$$x' = F(x)$$

where $F(\cdot)$ is the deep learning neural network. This problem can be represented as:

$$x' = \operatorname*{argmin}_{x} ||AF(x) - y||^2 + \lambda ||AF(x)||$$

where the $x'$ is sparse. The radar image $y$ is also sparse in some situation. We propose a method to learn the mapping function between the $y$ and $y'$. It can be considered as:

$$y' = F(y)$$

this problem can be represented as:

$$y' = \operatorname*{argmin}_{y} ||F(y) - y||^2 + \lambda ||F(y) - z||$$

where $z$ is image with the zeros value. Due to the $y'$ is similar to the $y$, we adapt convolutional neural network to learn the mapping function between the $y$ and $y'$. The loss function is considered as:

$$Loss = L2(F(y) - y) + \lambda L1(F(y))$$

where L2 is the L2-norm and L1 is the L1-norm.

B. Network Architecture

In this task, we adapt the network architecture similar to the paper [12]. The final network structure is shown in Fig.1.

As we can see that, this network has three layers. The first layer employs the convolution with size of 3x3, and the dimension of layer is set as 64. The second layer adopts the convolution with size of 3x3, and the dimension of layer is 64. In the last layer, the convolution kernel size is 3x3 and the dimension of layer is one. It represented the output image. In the training phase, ReLU is used as activation function, and all of the convolution operations keep the dimensions of images.

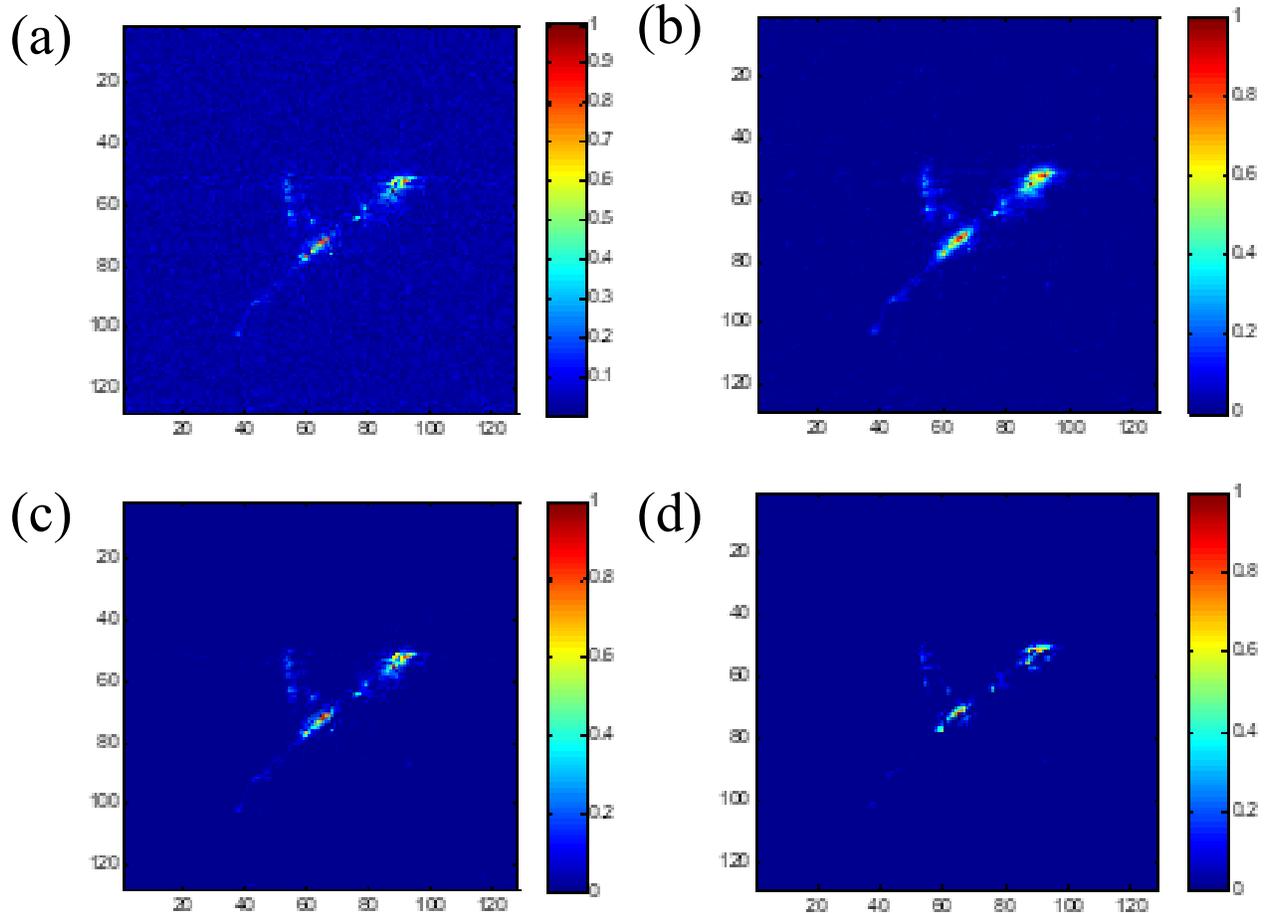

Fig.2 ISAR imaging result using the B-727 data set. (a) RD, (b) SS-DL-CS-Net with $\lambda = 0.1$, (c) SS-DL-CS-Net with $\lambda = 0.2$, (d) SS-DL-CS-Net with $\lambda = 0.3$

C. Training of network

To obtain the high quality radar image, the original radar image is used as the input of network. The original radar image and image with the zeros value is used as the reference of network. We use TensorFlow framework to implement our network, and the training parameter of the network is listed in Tab. 1.

Tab.1 Training parameters

| Number of training data | 1 |
|---|---|
| Gradient optimization | Adam algorithm |
| Learning rate | 0.0001 |
| Epoch | 100 |

## 3. EXPERIMENTS

A. Experimental Results On Measurement Data

We use the B-727 data to validate the SS-DL-CS-Net algorithm. The parameters of the data are shown as follows: carrier frequency: 9GHz, bandwidth: 150MHz, data size (range direction azimuth direction) : 128×128.

Fig. 2(a) is the ISAR imaging result using range-Doppler (RD) algorithm. As we can see that the target images generated by RD is very noisy. The Fig. 2(a) is used as the input of network called SS-DL-CS-Net. The Fig. 2(a) is also used as the reference of network as showed in Fig. 1. Fig. 2(b-d) shows the ISAR imaging results generated by SS-DL-CS-Net with $\lambda = 0.1$, $\lambda =$

0.2 and λ = 0.3. It can be found that as the increases of λ, ISAR imaging results with higher resolution and stronger noise suppression.

## 4. CONCLUSION

A novel enhanced radar imaging methods called SS-DL-CS-Net is proposed to obtain high quality radar image. We adapt it to ISAR image, and a clear ISAR image with high-resolution was obtained. With the development of convolutional neural network theory, we believe that CNN will bring more benefits to radar imaging technology in the future.